\documentclass[twocolumn,showpacs, preprintnumbers,amsmath,amssymb]{revtex4}
\usepackage{graphicx}
\usepackage{dcolumn}
\usepackage{bm}
\begin{document}
\title{ An $SU(2)_L\times U(1)_Y$ model with reflection
        symmetry  in view \\
        of recent neutrino experimental result}
       
\author{Ambar Ghosal}
\affiliation{
        Saha Institute of Nuclear Physics,
        1/AF Bidhannagar, Kolkata 700064, India}
        \thanks{E-mail address: ambar@anp.saha.ernet.in}
\date{\today}
\begin{abstract}
We demonstrate that
an $SU(2)_L\times U(1)_Y$ model with the same
particle content as
Standard Model (SM) and discrete reflection
symmetry between second and
third generation of leptons gives rise
to charged lepton and
neutrino mass matrices which can accommodate
the present solar, atmospheric,
WMAP neutrino experimental results.
The model predicts
the value of $|U_{13}|$ which could be tested
in neutrino factories and the value of
effective Majorana neutrino mass
$\langle m_{ee}\rangle$ comes out
at the lower end of the present
experimental limit.
Neutrino masses are generated through
dim=5 operators and
the scale of which is constrained by
the value of $\langle m_{ee}\rangle$.
If, in future neutrino
less double beta decay experiments namely, MOON, EXO , 
GENIUS etc. shifts 
the lower bound on
$\langle m_{ee}\rangle$ by one order the present model 
will fail to 
accommodate the solar neutrino mixing angle due to LMA 
solution. 
\end{abstract}
\maketitle
\narrowtext
\section{Introduction}
\noindent
Super-Kamiokande(SK) atmospheric neutrino experiment 
\cite{one}
has strengthened the conjecture of neutrino flavor oscillation 
as well as non-zero neutrino mass through the measurement of magnitude 
and angular distribution of $\nu_\mu$ flux produced in the atmosphere 
due to cosmic ray interactions and the data provide the following 
values as 
\cite{two} 
$1.5\times {10}^{-3} \rm{eV^2} \leq 
\Delta m^2_{atm}\leq 5.0\times {10}^{-3} \rm{eV^2}$, 
$\sin^22\theta_{atm}\ge 0.85$ at $99.73\%$ c.l.. 
Furthermore,  
the K2K reactor neutrino experimental
result 
\cite{three} 
is also in concordance with the interpretation
of atmospheric neutrino experimental result.
Taking into account the  
K2K reactor neutrino experimental result and the SK neutrino 
experimental result, the best fitted value obtained as 
\cite{best} 
$\Delta m^2_{atm}$= $2.7\times {10}^{-3} \rm{eV^2}$,
$\sin^22\theta_{atm}= 1.0$. 
Furthermore, it has been shown that 
\cite{four} 
global analysis 
of all solar neutrino experimental results, including SNO 
experiment 
\cite{five}, 
is in favor of Large Mixing Angle (LMA) 
MSW solution of solar $\nu_e$ deficit problem. This solution 
is also recently supported by the KamLAND reactor neutrino 
experimental result 
\cite{kam}.Combined analysis of KamLAND and all 
other solar neutrino 
experimental data restrict the parameter space of 
$\nu_e$ oscillation
\cite{othersol,bestsol} 
and there are slightly different ranges of $\Delta m^2_{\odot}$ 
and $\tan^2\theta_{\odot}$ have been found in the literature 
\cite{othersol} although the variation is not much. 
We consider 
for the present analysis the following ranges
\cite{bestsol} 
$5.32\times {10}^{-5} \rm{eV^2} \leq
\Delta m^2_{\odot}\leq 9.82\times {10}^{-5} \rm{eV^2}$,
$0.266\leq \tan^2\theta_{\odot}\leq 0.88$ at 3$\sigma$ level 
with the best-fitted value of 
$\Delta m^2_{\odot}$= $7.17\times {10}^{-5} \rm{eV^2}$,
$\tan^2\theta_{\odot}= 0.45$.      
Moreover, assuming $\theta_\odot$ as $\theta_{12}$ and 
 $\theta_{atm}$ as $\theta_{23}$, the CHOOZ experiment 
\cite{eight} 
has also put an upper bound on the 
$U_{e3}$ element of the lepton mixing matrix ( which 
is also known as MNSP matrix). It has been shown 
\cite{best,nine}
that combined three-neutrino oscillation analysis of 
solar, atmospheric and reactor data gives a lower bound on 
$U_{e3}$ as $|U_{e3}|\le 0.22$ at $99.73\%$ c.l. There are 
two other important experimental results, one of them, recently 
reported WMAP 
\cite{10} 
result on cosmic microwave background 
anisotropies. Combining analysis of 2dF Galaxy Redshift survey, 
CBI and ACBAR 
\cite{11}, 
WMAP has determined the amount of 
critical density contributed by relativistic neutrinos which 
in turn gives an upper bound on the total neutrino mass as 
$\Sigma m_i\le 0.71$ eV at $95\%$ c.l.
\cite{12}. 
Furthermor, 
neutrinoless double beta decay
($\beta\beta_{0\nu}$) experiment
\cite{13} 
has reported the bound on the effective Majorana 
neutrino mass 
(relaxing the uncertanity of the Nuclear Matrix elements 
up to $\pm 50\%$ and the contribution to this process due 
to particles other than Majorana neutrino is negligible 
\cite{14})
as 
$\langle m_{ee}\rangle= (0.05-0.84)$ eV at $95\%$ c.l. 
\cite{13}. 
Keeping all these 
constraints in view, it is very much interesting to 
find out an appropriate texture of lepton mass matrix 
which satisfy all those experimental results with 
appropriate mixing. There is much literature investigating
appropriate texture of neutrino mass matrix keeping 
charged lepton mass matrix diagonal. Another scenario, 
which considers both charged lepton and neutrino mass 
matrices are non-diagonal.       
\vskip .1in
\noindent
Recently, an investigation in this path has been done 
in Ref.[17], through the introduction of a discrete 
$Z_3$ symmetry within the framework of an SU(5) model 
to generate appropriate mixing in both quark and 
lepton sector. Neutrino masses are generated in this 
model through see-saw mechanism and the model 
gives rise to appropriate mixing which reconciles 
with the neutrino experimental results. In the present 
work, we consider a different texture of charged lepton 
and neutrino mass matrices within the framework of an 
$SU(2)_L\times U(1)_Y$ model with a reflection symmetry 
and keeping the particle content same as
Standard Model (SM).
Neutrino masses are generated through dim=5 operators 
due to explicit violation of Lepton number. In addition, 
we consider a reflection symmetry between second and third 
generation of leptons. We have considered both charged 
lepton and neutrino mass matrices non-diagonal. Apart
from the phase factor, the structure of 
neutrino mass matrix is same as the texture 
of neutrino mass matrix  
presneted in 
Ref.[18]. 
The parameter space of the model admits solar, 
atmospheric and WMAP experimental results and predicts 
the value of $\theta_{13}$ well below the CHOOZ experimental 
result, the value of which may be possible to check in 
future neutrino factories 
\cite{20}. 
The present model also 
predicts the value of effective Majorana neutrino 
mass $\langle m_{ee} \rangle$
which is at the lower end of the experimental
limit. If the future
${\beta\beta}_{0\nu}$ experiments, such as,
MOON, EXO or GENIUS
experiments ( which could be possibly put
more stringent bound on $\langle m_{ee}\rangle$), 
pull the lower bound on 
$\langle m_{ee}\rangle$ by one order 
the present model will be ruled out. The 
effective scale of the present model 
is also constrained by the 
value of $\langle m_{ee}\rangle$
comply with the unitarity bound of the
Yukawa couplings. 
In section II, we present the model and the 
neutrino and charged lepton mass matrices.  
Neutrino phenomenology is discussed in Section 
III. Section IV contains summary of the present 
work.       
\section{Proposed Model}
\noindent
We consider an $SU(2)_L\times U(1)_Y$ model with particle 
content same as Standard Model. We consider the following 
reflection symmetry in the lepton sector 
$$
l_{2L}\leftrightarrow l_{3L}, 
\mu_R\leftrightarrow \tau_R \, .
\eqno(2.1)
$$
\noindent
The most general discrete symmetry invariant lepton-Higgs
Yukawa interaction in the present model which gives rise to
charged lepton and neutrino mass is given by
$$
L_Y^E =
[g_{11}{\overline{l_{1L}}}e_R +
 g_{12}{\overline{l_{1L}}}(\mu_R + \tau_R) +
$$
$$
 g_{21}({\overline{l_{2L}}} + {\overline{l_{3L}}})e_R
 +g_{22}({\overline{l_{2L}}}\mu_R + {\overline{l_{3L}}}\tau_R)
$$
$$
 +
 g_{23}({\overline{l_{2L}}}\tau_R + {\overline{l_{3L}}}\mu_R)]
 \phi 
+ h. c.
 \eqno(2.2)
 $$
 $$
 L_Y^\nu
 =
 [f_1 l_{1L}l_{1L}
 + f_2 l_{1L}(l_{2L}+l_{3L}) +
 f_3(l_{2L}l_{2L}
$$
$$
+l_{3L}l_{3L}
+l_{2L}l_{3L})]{\phi\phi/M}
 \eqno(2.3)
 $$
\noindent
where $M$ is an additional scale of the theory apart from
the electroweak symmetry breaking scale. Substituting
VEV of the Higgs field $\phi$,
$\langle\phi^0\rangle = v$,
we obtain the charged
lepton and neutrino mass matrices as
$$
M_E = \begin{pmatrix}a&b&b\cr
                     c&d&e\cr
                     c&e&d
\end{pmatrix}
\eqno(2.4)
$$
where
$$
a = g_{11}v , b = g_{12}v , c = g_{21}v,
$$
$$
d = g_{22}v, e = g_{23}v
\eqno(2.5)
$$
$$
M_\nu =\begin{pmatrix}p&q&q\cr
               q&r&r\cr
               q&r&r
\end{pmatrix}
\eqno(2.6)
$$
\noindent
where
$$
p = f_1\lambda,\,\, q = f_2\lambda,\,\,
r = f_3\lambda,\,\,
\lambda = v^2/M
\eqno(2.7)
$$
\noindent
To diagonalize the charged lepton and neutrino
mass matrix, we consider an orthogonal matrix,
$O_{\alpha}$, ($\alpha = e, \nu$) as
$$
O_{\alpha} = \begin{pmatrix}c_{12}^{\alpha}c_{31}^{\alpha}&
                      c_{31}^{\alpha}s_{12}^{\alpha}&
                      s_{31}^{\alpha}\cr
&&\cr
-s_{12}^{\alpha}c_{23}^{\alpha}&c_{12}^{\alpha}c_{23}^{\alpha}&
s_{23}^{\alpha}c_{31}^{\alpha}\cr
-s_{23}^{\alpha}s_{31}^{\alpha}c_{12}^{\alpha}&
-s_{23}^{\alpha}s_{31}^{\alpha}s_{12}^{\alpha}&&\cr
&&\cr
s_{23}^{\alpha}s_{12}^{\alpha}&-s_{23}^{\alpha}c_{12}^{\alpha}&
c_{23}^{\alpha}c_{31}^{\alpha}\cr
-c_{23}^{\alpha}s_{31}^{\alpha}c_{12}^{\alpha}&
-s_{31}^{\alpha}s_{12}^{\alpha}c_{23}^{\alpha}&&
\end{pmatrix}
\eqno(2.8)
$$
where $O_e$ and $O_\nu$ are utilised to diagonal
$M_E$ and $M_\nu$, respectively.
\vskip .1in
\noindent
By considering $M_E$ is real, we  diagonalize
the charged lepton mass matrix as
$$
O_{e}^TM_E M_E^T O_e = {\rm diag}(m_1^2, m_2^2, m_3^2)
\eqno(2.9)
$$
\noindent
and we obtain the three eigenvalues as
$$
m_1^2 = \frac{{(X+W+Z)-\sqrt{{(X-W-Z)}^2 + 8Y^2}}}{2}
$$
$$
m_2^2 = \frac{{(X+W+Z)+\sqrt{{(X-W-Z)}^2 + 8Y^2}}}{2}
$$
$$
m_3^2 = Z-W
\eqno(2.10)
$$
\noindent
and the three mixing angles are
$$
\theta_{31}^e= 0,
\theta_{23}^e = -\pi/4,
$$
$$
\tan^2\theta_{12}^e = \frac{(X-m_1^2)}{(m_2^2-X)}
\eqno(2.11)
$$
where
$$
X = a^2 + 2 b^2,
Y = ac +bd +be,
Z = c^2 +d^2 + e^2,$$
$$
W = c^2 +2de
\eqno(2.12)
$$
\vskip .1in
\noindent
The neutrino mass matrix $M_\nu$ is diagonalized as
$$
(P_\nu^\dagger O_\nu)^\dagger D_\nu (P_\nu O_\nu)
= M_\nu
\eqno(2.13)
$$
where
$$
D_{\nu} = {\rm{diag}}(m_1^\nu, m_2^\nu, m_3^\nu)
\eqno(2.14)
$$

$$
P_\nu = \begin{pmatrix}1&0&0\cr
                 0&e^{i\delta_2^\nu}&0\cr
                 0&0&e^{i\delta_3^\nu}
\end{pmatrix}
\eqno(2.15)
$$
and the three neutrino masses come out as
$$
m_1^\nu = \frac{(p+2r) - \sqrt{p^2+8q^2}}{2}
$$
$$
m_2^\nu = \frac{(p+2r) + \sqrt{p^2+8q^2}}{2}
$$
$$
m_3^\nu = 0
\eqno(2.16)
$$
with the three mixing angles are given by
$$
\theta_{31}^\nu = 0,
\theta_{23}^\nu = -\pi/4,
$$
$$
\tan^2\theta_{12}^\nu = (p - m_1)/(m_2 - p)
\eqno(2.17)
$$
where $m_1$, $m_2$ are the
charged lepton masses given
in eqn.(2.10).
It is to be noted that, the symmetry proposed in the model
given in eqn.(2.1) constraints the relationship between
$\delta_2^\nu$ and $\delta_3^\nu$ as
$\delta_2^\nu$ = -$\delta_3^\nu$. More precisely, this
can be understand if we consider
$l_{2L}\rightarrow e^{i\delta_2^\nu}l_{3L}$,
$l_{3L}\rightarrow e^{i\delta_3^\nu}l_{2L}$
which necessarilly leads
$\delta_2^\nu = -\delta_3^\nu $. However, such transformation
also leads charged lepton mass matrix complex. Further
demanding the symmetry
$\mu_R\rightarrow  e^{i\alpha_1}\tau_R$, 
$\tau_R\rightarrow  e^{i\alpha_2}\mu_R$ (and as previous
$\alpha_1 = -\alpha_2$) will give rise to a phase matrix
$P_e$ which is similar in structure to $P_\nu$. Both
$P_e$ and $P_\nu$ will appear in $U^{\rm MNSP}$ matrix
(Maki - Nakagawa - Sakata - Pontecorvo)
and
the product $P_e^\dagger P_\nu = P$ = $diag(1, e^{i\delta_2},
e^{i\delta_3})$  has the same structure as $P_e$ or
$P_\nu$ with the constraint relation
$\delta_2 = -\delta_3$. Furthermore if we consider Dirac
phase in $M_E$ the structure of $U^{\rm MNSP}$ will not
change.
Thus the MNSP 
mixing matrix appear in the charged current
interaction as
$$
U^{\rm{MNSP}} = O_e^T P O_\nu
$$
$$
=
\begin{pmatrix}c^ec^\nu + \rho s^e s^\nu&
         c^es^\nu - \rho s^e c^\nu&         
         -\sigma s^e\cr
         s^ec^\nu - \rho c^e s^\nu&
         s^es^\nu +\rho c^e c^\nu&         
         \sigma c^e\cr
         -\sigma s^\nu&\sigma c^\nu& \rho
\end{pmatrix}
\eqno(2.18)
$$
where
$$
c^{e,\nu} = c_{12}^{e,\nu},\,\, 
s^{e,\nu} = s_{12}^{e,\nu},
$$
$$
\rho = \frac{(e^{i\delta_2}+e^{i\delta_3)}}{2},
\sigma = \frac{(e^{i\delta_3}-e^{i\delta_2)}}{2} \,\, .
\eqno(2.19)
$$
\vskip .1in
\noindent
Here, we have used the notation of Ref.[17]. It is to be noted 
that if $\delta_2 = \delta_3 = 0$, the mixing matrix becomes 
unphysical,
thus, if we ignore any complex nature of $M_\nu$
and/or $M_E$,
the present model will not produce any viable
$U^{MNSP}$ matrix.  
\section{Neutrino Phenomenology}
\noindent
Let us discuss neutrino phenomenology of the present model. 
First of all, we have fix the value of $m_2^\nu$ by the 
required mass-squared difference needed to explain 
atmospheric neutrino problem as
$$
\Delta m^2_{atm} = \Delta m^2_{23} = {(m_2^\nu)}^2
= 2.7\times {10}^{-3}{\rm eV^2}
\eqno(3.1)
$$
\noindent
where we have used the best-fitted value of 
$\Delta m^2_{atm}$ and $m_2^\nu$ comes out as 
$m_2^\nu = 0.0519$ eV. Next, we fix the value of 
$m_\nu^1$ with the solar neutrino experimental result as 
$\Delta m^2_{\odot}$ = $\Delta m^2_{21}$ =
$ 7.1\times {10}^{-5}{\rm eV^2}$ and the value of 
$m_1^\nu$ becomes
$m_1^\nu$ = 0.0512 eV. The three mixing angles
required to explain the solar, atmospheric and CHOOZ neutrino 
experimental results are given by
$$
\sin^22\theta_{atm} =  4{|U_{23}|}^2{|U_{33}|}^2$$
                   $$ =  4{|\rho|}^2{|\sigma|}^2c_e^2$$
                   $$ =  \frac{(m_2^2-X)}{(m_2^2-m_1^2)}
                          \sin^2(\delta_2-\delta_3) 
\eqno(3.2)
$$
where $m_1$, $m_2$ are the masses of charged leptons 
given in eqn.(2.10). Similarly, 
$$
{|U_{13}|}^2 = \frac{1}{2}\frac{(X-m_1^2)}{(m_2^2-m_1^2)}
\eqno(3.3)
$$
and 
$$
\sin^22\theta_{\rm{CHOOZ}} = 4{|U_{13}|}^2(1 - {|U_{13}|}^2)
\eqno(3.4)
$$
$$
\sin^22\theta_{\odot} = 4 {|U_{11}|}^2 {|U_{12}|}^2$$
$$ = 
  4[c_e^2c_\nu^2 + \frac{s_e^2s_\nu^2}{2}
  + \sqrt{2}c_e c_\nu s_e s_\nu 
\cos\frac{\delta_2+\delta_3}{2}]\times
$$
$$
 [c_e^2c_\nu^2 + \frac{s_e^2c_\nu^2}{2}
   - \sqrt{2}c_e c_\nu^2 s_e \cos\frac{\delta_2+\delta_3}{2}]
\eqno(3.5)
$$ 
Using the value of $m_1$,$m_2$ as
$m_1$ = 0.4868 MeV, $m_2$ = 105.7513 MeV,
we now fix the value of $X$ from eqn.(3.2).
It is obvious from eqn.(3.2) that $\delta_2\neq \delta_3$ and 
$X\neq m_2^2$ to get 
non-zero atmospheric mixing angle.
However, if $X = m_1^2$ and 
$\delta_2 = -\delta_3 = \pi/4$ the atmospheric 
mixing angle becomes $\sin^22\theta_{atm} = 1.0$ and in that case 
$|U_{13}|$ = 0 and the solar neutrino mixing angle also shoots up 
to maximal value which is an ideal case for bi-maximal neutrino 
mixing. To find out a  realistic parameter space, the rest of 
the work  
we consider $\delta_2 = -\delta_3 = \pi/4$ 
and from numerical estimation we find the value of $X$ as 
$X = m_1m_2/100 $ and with such value of $X$ we get 
$$
\sin^22\theta_{atm} = 0.99\,\,\, \rm{and}$$
$$
\sin^22\theta_{13}= 4.99\times {10}^{-5},
{|U_{13}|}^2\simeq 1.25\times {10}^{-5}
\eqno(3.6)
$$
Such small value of $\theta_{13}$ or $U_{13}$ 
may be probed in future 
neutrino factories with superbeam facilities 
\cite{20}. 
The      
solar neutrino mixing angle comes out as 
$\sin^22\theta_{\odot}\simeq 0.85$ for the choice 
of $p = 0.0516$ eV. The parameter space is very sensitive to 
the parameter $p$ and $p$ can be varied upto 
$p< 0.0515$ eV so that $\sin^22\theta_{\odot}<1$. 
\vskip .1in
\noindent
The parameter $p$ also determines the value of 
effective Majorana neutrino mass
$\langle m_{ee}\rangle$ = $p$
 and the value obtained 
marginally satisfies the bound obtained 
from recent $\beta\beta_{0\nu}$ decay. Thus, if the lower 
bound of $\langle m_{ee}\rangle$
shifts by one order, the present model
will fail to accommodate the solar neutrino 
mixing angle. Furthermore, the value of $p$ also detremines the 
scale $M$ of the present model by considering perturbative 
unitarity bound on the Yukawa coupling $f_1$ as 
$f_1^2/4\pi\leq 1$. Thus, the additional scale $M$ of the theory 
can be written as 
$$
M = \frac{f_1{<\phi>}^2}{p}\leq \frac{2\sqrt{\pi}{<\phi>}^2}{p}  
\eqno(3.6)
$$
and for $\langle\phi\rangle$ $\simeq$ 200 GeV , we get 
$M\leq{10}^{13}$ GeV. Finally, in the present model, sum of the 
three neutrino mass comes out
as $\Sigma m_i$ = 0.1031 eV
which is also far below than the bound obtained from WMAP 
experimental result. 
\section{Summary}
\vskip .1in
\noindent
We demonstrate that an $SU(2)_L\times U(1)_Y$ model with particle 
content same as SM and with a discrete reflection symmetry between 
second and third generation of leptons give rise to an 
appropriate texture of charged lepton and neutrino mass matrices 
which can accommodate the present atmospheric, solar, CHOOZ, WMAP and 
${\beta\beta}_{0\nu}$ decay experimental results. Neutrino masses 
are generated through explicit lepton number violation due to 
dim = 5 mass terms and the charged lepton masses are generated 
according to SM. Apart from electroweak symmetry breaking scale, 
the present model contains an additional scale $M$, which is constrained 
as $M\leq {10}^{13}$ GeV due to  
the ${\beta\beta}_{0\nu}$ decay experimental result and if the present 
lower bound of the ${\beta\beta}_{0\nu}$ decay result shifts by one order
in future ${\beta\beta}_{0\nu}$ experiments, such as, MOON, EXO, GENIUS, 
the present model will fail to accommodate the solar neutrino 
mixing angle due to LMA solution recently favoured by the SNO and 
KamLAND experimental results. It is worthwhile to investigate application 
of such symmetry in the context of other models where neutrino masses 
are generated through see-saw mechanism or by radiative way.    

\begin{acknowledgments}
The author acknowledges Abhijit Bandyopadhyay 
for many helpful discussions.
\end{acknowledgments}
 
\end{document}